# Super- and subradiance by entangled free particles

Aviv Karnieli[1,*], Nicholas Rivera[2], Ady Arie[3] and Ido Kaminer[4]

[1]Raymond and Beverly Sackler School of Physics and Astronomy, Tel Aviv University, Ramat Aviv 69978, Tel Aviv, Israel
[2]Department of Physics, Massachusetts Institute of Technology, Cambridge, Massachusetts 02139, USA
[3]School of Electrical Engineering, Fleischman Faculty of Engineering, Tel Aviv University, Tel Aviv 69978, Israel
[4]Department of Electrical Engineering, Technion–Israel Institute of Technology, Haifa 32000, Israel

*corresponding authors: kaminer@technion.ac.il; avivkarnieli@tauex.tau.ac.il

When multiple quantum emitters radiate, their emission rate may be enhanced or suppressed due to collective interference in a process known as super- or subradiance. Such processes are well-known to occur also in light emission by free charged particles. To date, all experimental and theoretical studies of super- and subradiance in these systems involved the classical correlations between the emitters. However, dependence on quantum correlations, such as entanglement between different emitting particles, has not been studied. Recent advances in coherent-shaping of free-electron wavefunctions motivate the investigation of such quantum regimes of super- and subradiance. In this Letter, we show how a pair of coincident path-entangled electrons can demonstrate either super- or subradiant light emission, depending on the two-particle wavefunction. By choosing different free-electron Bell-states, the spectrum and emission pattern of the light can be reshaped, in a manner that cannot be accounted for by a classical mixed state. We show these results for light emission in any optical medium, and discuss their generalization to many-body quantum states. Our findings suggest that light emission can be sensitive to the explicit quantum state of the emitting matter wave, and possibly serve as a non-destructive measurement scheme for measuring the quantum state of many-body systems.

## Introduction

Quantum electrodynamics (QED)[1] has proven itself to be one of the most accurate physical theories to date, and since its discovery has managed to shed light over the fundamental aspects of light-matter interactions. One surprising implication of the quantum theory of light and matter is the enhancement (or suppression) of the spontaneous emission rate of a system of $N$ emitters above (below) the rate of $N$ independent emitters[2,3]. These superradiance (subradiance) effects have been thoroughly studied for systems of bound electrons[4–9].

Such phenomena also occur for classical beams of free charged particle: superradiant emission is known to scale quadratically with the particle number[10–14], which constitutes the feedback and gain mechanisms of free electron lasers (FELs)[11,12,15]. In addition, subradiant emission below shot noise was demonstrated, exploiting the Coulomb repulsion between particles[16–18]. Controlling the shape, spectrum, and photon statistics of spontaneous emission by free-electrons has gained recent attention[19–23] for its potential use as an efficient light source at otherwise inaccessible wavelengths, and for its novel prospects for quantum optics[24–28].

In contrast to the super- and subradiance from bound-electrons case that necessitates a quantum theory, super- and subradiance from free electrons (and other charged particles) was described classically in all experimental regimes so far[10–14,29]. In classical physics, waves interfere coherently when they are generated from different point particles that are perfectly correlated with each other (a *bunch*, see Fig. 1a)[30]. The bunching of free charged particles leading to spontaneous superradiant (or subradiant) emission is seen in two scenarios: (1) when the bunch size is smaller than the wavelength and (2), when the spatial modulation of the electron density is an integer multiple of the emission wavelength. In contrast, classical superradiance and subradiance are diminished for randomly positioned particles in a bunch much larger than the emitted wavelength.

To date, even the quantum description of superradiance by multiple free charged particles[31–35] recovered the same predictions as the classical description [endnote: The known differences between the classical and the quantum descriptions are not related to the number of particles. Rather, even at the level of a single particle emitter, there exist small quantum corrections due to quantum recoil, i.e., the quantized nature of the emitted light[36]]. The quantum description involves position-momentum uncertainty of each particle in the multiparticle wavefunction. The wavefunction creates a spatio-temporal distribution of the matter wave, which can create coherent interference of light. This interference matches the classical case, in which super- and subradiance depend on a continuous charge distribution. In other words, the quantum and classical descriptions provide the same predictions. In the current paradigm the emission pattern and spectrum are determined solely by the structure factor – or the Fourier transform of the charge density – whether it is described as a classical bunch or as a quantum multiparticle wavefunction.

Here, we ask a different question: can *quantum correlations* in the multiparticle wavefunction control and shape light emission? Are there special multiparticle states for which super- and subradiance behave differently than in the classical case? We answer in the affirmative – quantum correlations between multiple emitters shape light emission by creating a new effect of quantum super- and sub-radiance. To show this, we develop the general theory of spontaneous emission by a multiparticle wavefunction of free charged particles. For the experimentally-common case in which the particles are free electrons, the process of light emission is also called coherent cathodoluminescence[37] (CCL). We adopt the acronym CCL below, while keeping in mind that the predictions apply to other charged particles.

To exemplify the general concept of quantum super- and subradiance for multiparticle CCL, we present results for the concrete case of light emission by two path-entangled free electrons. We consider each electron as having a delocalized wavepacket larger than the emitted wavelength. This case therefore naively corresponds to incoherent emission[38] in both the classical and the quantum descriptions. Surprisingly, we find that the emitted light intensity directly depends on the quantum phase angle of the two-electron Bell-state, and that both super- and subradiant light emission can be obtained for different quantum states. This phenomenon has no classical analogue, and we discuss how to distinguish it from cases of "classical" super- and subradiance. This finding, therefore, constitutes a yet-unexplored regime of quantum super- and subradiance by free charged particles. Our findings have implications on the emerging field of quantum optics of free electron light sources[24,25,39], suggesting that photoemission can unveil information on the quantum state of multi-particle emitters[38].

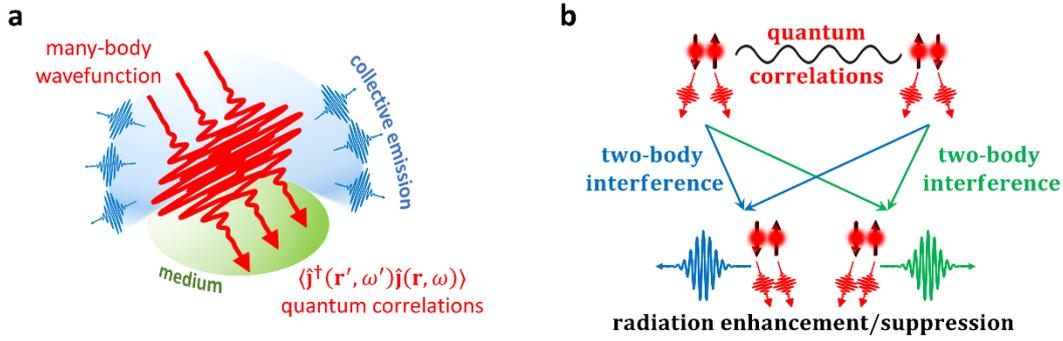

**Fig. 1: Super- and subradiance from quantum-correlated free charged particles.** (a) We can define a quantum current operator $\hat{\mathbf{j}}(\mathbf{r}, t)$ that is associated with the emission of light quanta by multiple quantum charged particles in a general optical environment. The current operator can be used to find the collective (super- or subradiant) emission by calculating current-current correlations. (b) Exemplifying the general concept, when a pair of quantum-correlated particles emits radiation, the quantum interference between the transition amplitudes can lead to enhancement or suppression of the emitted light intensity.

**CCL by free charged particles.** To illustrate our findings, we consider CCL by free electrons in a general optical environment described by a dyadic Green function[40,41] $\mathbf{G}(\mathbf{r}, \mathbf{r}', \omega)$. The initial state of the electron-radiation field is described by a density matrix $\boldsymbol{\rho}_\mathrm{i} = \boldsymbol{\rho}_\mathrm{e} \otimes |0\rangle\langle 0|$, where $\boldsymbol{\rho}_\mathrm{e}$ denotes the initial electron density matrix. The interaction is governed by the Dirac Hamiltonian: $H_\mathrm{int} = ec\boldsymbol{\alpha} \cdot \mathbf{A}$, where $e$ is the electron charge, $c$ is the speed of light, $\boldsymbol{\alpha}^i = \boldsymbol{\gamma}^0\boldsymbol{\gamma}^i$ are the Dirac matrices, and $\mathbf{A}$ is the electromagnetic vector potential operator. We use first-order time-dependent perturbation theory to find the final quantum state of the system, $\boldsymbol{\rho}_\mathrm{f}$.

In CCL experiments, only the radiation field is measured and so we calculate its reduced density matrix, $\boldsymbol{\rho}_\mathrm{ph} = \mathrm{tr}_\mathrm{e}\{\boldsymbol{\rho}_\mathrm{f}\}$, with $\mathrm{tr}_\mathrm{e}$ denoting the partial trace over the (multi-)electron state. We calculate the power spectrum of the emitted light measured in the far field at a distance $r$ from the source and at direction $\hat{\mathbf{n}}$ (see Supplementary Material Sections S1-S3 for derivation):

$$\frac{d^2 P}{d\Omega d\omega} = 2r^2 \epsilon_0 c \omega^2 \mu_0^2 \int d^3\mathbf{R} d^3\mathbf{R}' \, \mathrm{Tr} \mathbf{G}^\dagger(r\hat{\mathbf{n}}, \mathbf{R}', \omega) \mathbf{G}(r\hat{\mathbf{n}}, \mathbf{R}, \omega) \langle \mathbf{j}^\dagger(\mathbf{R}', \omega) \mathbf{j}(\mathbf{R}, \omega) \rangle_\mathrm{e} \qquad (1)$$

In Eq. 1, Tr denotes a matrix trace $\text{Tr}\,\mathbf{E}^\dagger\mathbf{E} = \sum_\alpha E_\alpha^\dagger E_\alpha$ over the electric field polarization. The quantity $\langle \mathbf{j}^\dagger(\mathbf{r}',\omega')\mathbf{j}(\mathbf{r},\omega)\rangle_\text{e} = \text{tr}\{\boldsymbol{\rho}_\text{e}\mathbf{j}^\dagger\mathbf{j}\}$ is the expectation value, with respect to the *electronic initial state*, of the correlations in the current density operator $\mathbf{j}(\mathbf{r},t) = ec\Psi^\dagger\boldsymbol{\alpha}\Psi$, where $\Psi(\mathbf{r},t)$ is the electron spinor field operator described in second quantization.

We now make the two following simplifications: (i) the particles propagate as wavepackets with a well-defined carrier velocity $\mathbf{v}_0$ (equivalent to the paraxial approximation, where the electron dispersion is linearized); (ii) photon-induced recoil associated with the momentum $\hbar q$ are much smaller than electron momenta $p_\text{e}$. These assumptions are applicable to a vast number of effects, including all cases in which the emitter is relativistic, all free-electron nanophotonic light sources, and all free-electron sources in the microwave and radio frequency ranges[20,22,37,42,43]. The current correlations appearing in Eq. 1 can then be written as (see Supplementary Material Section S2 for derivation)

$$\langle \mathbf{j}(\mathbf{x}')\mathbf{j}(\mathbf{x})\rangle = e^2 \mathbf{v}_0 \mathbf{v}_0 \left[ G_\text{e}^{(2)}(\mathbf{x}',\mathbf{x}) + \delta(\mathbf{x}-\mathbf{x}') G_\text{e}^{(1)}(\mathbf{x},\mathbf{x}) \right], \qquad (2)$$

where $\mathbf{x} = \mathbf{r} - \mathbf{v}_0 t$ and $\mathbf{x}' = \mathbf{r}' - \mathbf{v}_0 t'$. In Eq. (2), we define the first- and second-order correlation functions of the emitter $G_\text{e}^{(1)}(\mathbf{x}',\mathbf{x}) = \sum_\sigma \text{tr}\{\boldsymbol{\rho}_\text{e}\psi_\sigma^\dagger(\mathbf{x}')\psi_\sigma(\mathbf{x})\}$ and $G_\text{e}^{(2)}(\mathbf{x}',\mathbf{x}) = \sum_{\sigma'}\sum_\sigma \text{tr}\{\boldsymbol{\rho}_\text{e}\psi_{\sigma'}^\dagger(\mathbf{x}')\psi_\sigma^\dagger(\mathbf{x})\psi_\sigma(\mathbf{x})\psi_{\sigma'}(\mathbf{x}')\}$, respectively, where $\psi_\sigma(\mathbf{x})$ are position-space annihilation operators corresponding to the particle spin components $\sigma = \uparrow,\downarrow$. Eq. 2 is valid for particles with both fermionic and bosonic statistics. The current correlation of Eq. (2) comprises two terms: a pair correlation term proportional to $G_\text{e}^{(2)}(\mathbf{x}',\mathbf{x})$, giving rise to coherent radiation when substituted into Eq. (1); and a term proportional to the probability density $G_\text{e}^{(1)}(\mathbf{x},\mathbf{x})$, contributing incoherent radiation[38].

**Quantum super- and subradiance by free charged particles.** The quantum interference of the multiparticle wavefunction with itself can leave an imprint on the spontaneously-emitted light. Such quantum features are sensitive to the specific quantum state the system was prepared in and cannot be accounted for by classical electromagnetism. This phenomenon originates from the second-order correlation between the emitting particles, $G_\text{e}^{(2)}(\mathbf{x},\mathbf{x}')$ in Eq. (2), which determines the emission pattern in Eq. (1). Without loss of generality, in what follows we consider the decomposition

$$G_\text{e}^{(2)}(\mathbf{x},\mathbf{x}') = G_\text{e}^{(1)}(\mathbf{x},\mathbf{x}) G_\text{e}^{(1)}(\mathbf{x}',\mathbf{x}') g_\text{e}^{(2)}(\mathbf{x}-\mathbf{x}'). \qquad (3)$$

where $g_\text{e}^{(2)}(\mathbf{x}-\mathbf{x}')$ now stands for the normalized second-order correlation function of the emitting particles. It is worth mentioning that from Eqs. (1-3), the known semiclassical cases of superradiance could be recovered. For example, we get as special cases both the radiation from subwavelength-bunched beams[10], and the radiation from regularly-spaced bunches (as in the bunching effect in FELs[15]). In such cases, the spatio-temporal shape of the multiparticle wavefunction determines the power spectrum. Other second-order classical correlations known to lead to subradiance (e.g., as resulting from Coulomb interactions[16–18]) could be modelled by an appropriate choice of $g_\text{e}^{(2)}(\mathbf{x}-\mathbf{x}')$.

In the following, however, we shall focus on the case where $g_\text{e}^{(2)}(\mathbf{x}-\mathbf{x}')$ is determined by *quantum correlations* rather than classical correlations. As a proof-of-concept, we consider a two-electron state prepared from two identical, delocalized wavepackets $\varphi(\mathbf{x})$ having two possible carrier momenta: $\mathbf{k},\mathbf{k}'$ and different spins $\uparrow,\downarrow$, as can be created in a Stern–Gerlach-type experiment that splits electrons by spin into two directions of motion. Note that the spin degree of freedom, while usually insignificant for first-order spontaneous emission, becomes important here for the preparation of an entangled free-electron state. The electron pair traverses coincidentally through a general optical medium and spontaneously emits CCL radiation, as illustrated in Fig. (5a-b). We assume that the wavefunction momentum uncertainty around $\mathbf{k}, \mathbf{k}'$ is smaller than the difference $|\mathbf{k} - \mathbf{k}'|$, ensuring that the two states do not overlap in momentum space. Further, we assume that spatio-temporal walk-off (or group velocity mismatch) between the two wavepackets is negligible. This can be readily assured for electrons whenever $\mathbf{k} - \mathbf{k}'$ is of the order of optical momenta, and the wavefunction spatial extent in the respective dimension along $\mathbf{k} - \mathbf{k}'$ is larger than the optical wavelength.

Fig. 2 compares the cases of classical and quantum super- and subradiance. We illustrate the classical case in Fig. 2a and the quantum case in Fig. 2b. For the classical case, we analyze *probabilistic* correlations between the two electrons. If one is found in state $\mathbf{k} \uparrow$, it is correlated with the other to be found in $\mathbf{k}' \downarrow$, and vice versa. This corresponds to the mixed state $\rho_e = 1/2|\mathbf{k}_\uparrow \mathbf{k}'_\downarrow\rangle\langle \mathbf{k}_\uparrow \mathbf{k}'_\downarrow| + 1/2|\mathbf{k}_\downarrow \mathbf{k}'_\uparrow\rangle\langle \mathbf{k}_\downarrow \mathbf{k}'_\uparrow|$. From Eqs. (1-3), one may calculate the power spectrum $d^2P/d\Omega d\omega = \hbar\omega\Gamma(\hat{\mathbf{n}}, \omega)$ of the classically-correlated state, where $\Gamma(\hat{\mathbf{n}}, \omega)$ is the emission rate per unit time per unit frequency. For the classical case, $G_e^{(1)} = 2|\varphi(\mathbf{x})|^2$, and $g_e^{(2)}(\mathbf{x} - \mathbf{x}') = 1/2$. We denote the resulting classical emission rate by $\Gamma_c(\hat{\mathbf{n}}, \omega)$ and the single-particle emission by $\Gamma_0$. In the classical case, we find incoherent emission, i.e. $\Gamma_c = N\Gamma_0$ (with $N = 2$ in our case), for wavelengths smaller than the extent of the wavepacket $|\varphi(\mathbf{x})|^2$. We also find the expected classical superradiance, i.e. $\Gamma_c = N^2\Gamma_0$, for wavelengths larger than the extent of the wavepacket. See Fig. 2d for an example.

In the quantum case, we may consider a *fundamentally* different correlation, e.g., an entanglement between an electron pair prepared in a path-entangled Bell-state: $|\Psi\rangle = (|\mathbf{k}_\uparrow; \mathbf{k}'_\downarrow\rangle + e^{i\zeta}|\mathbf{k}_\downarrow; \mathbf{k}'_\uparrow\rangle)/\sqrt{2}$, where $\zeta$ is a phase angle. It is readily shown that in this case, we have again $G_e^{(1)} = 2|\varphi(\mathbf{x})|^2$, whereas

$$g_e^{(2)}(\mathbf{x} - \mathbf{x}') = \frac{1}{2}(1 - \cos\zeta \cos[\Delta\mathbf{k} \cdot (\mathbf{x} - \mathbf{x}')]). \tag{4}$$

Unlike the classical case of $g_e^{(2)} = 1/2$, the quantum case in Eq. (4) depends explicitly on the quantum phase angle $\zeta$ of the electron Bell-state, and also depends on $\Delta\mathbf{k} = \mathbf{k} - \mathbf{k}'$. Calculating the quantum emission rate $\Gamma_q$ from Eqs. (1-3) and (4) gives the general result

$$\Gamma_q = \Gamma_c - \cos\zeta\, \Gamma_{\Delta\mathbf{k}}, \tag{5}$$

where $\Gamma_{\Delta\mathbf{k}}$ is a term resulting from the momentum difference $\Delta\mathbf{k}$ of the two modes $\mathbf{k}$ and $\mathbf{k}'$. This term can influence the radiated spectrum by shifting the long wavelength superradiance peak in momentum space to shorter wavelengths (as shown below). This property, together with the control over the quantum phase angle $\zeta$, allows for selective **enhancement or suppression of the emission rate at wavelengths that exhibit no super- or subradiance in the classical picture**. i.e., we find a peak (or dip) in the emission intensity that cannot be explained by spatial modulation (bunching) of the density cloud $G_e^{(1)} = 2|\varphi(\mathbf{x})|^2$ (the two chosen wavepacket modes $e^{i\mathbf{k}\cdot\mathbf{x}}\varphi(\mathbf{x})$ and $e^{i\mathbf{k}'\cdot\mathbf{x}}\varphi(\mathbf{x})$ differ only by phase and are not modulated in amplitude). Therefore, Eq. (5) introduces a *quantum* radiation effect that is sensitive to the quantum correlation between electrons and shows how they induce strong enhancement or suppression of spontaneous emission.

Note that the $\cos\zeta$ term of Eq. (5) – the quantum interference in superradiance and subradiance – happens *independently* of any temporal delay effects between electrons, as previously analyzed in the literature[31,32]. In fact, the radiation phenomenon we described does not rely on the localization of the wavefunction to dimensions smaller than the emitted wavelengths, nor on temporal separation between electrons (see the appendix and Supplementary Material for further derivations and comparison). For this reason, the example we provide below focuses on a case of two delocalized electron wavepackets arriving at the sample simultaneously.

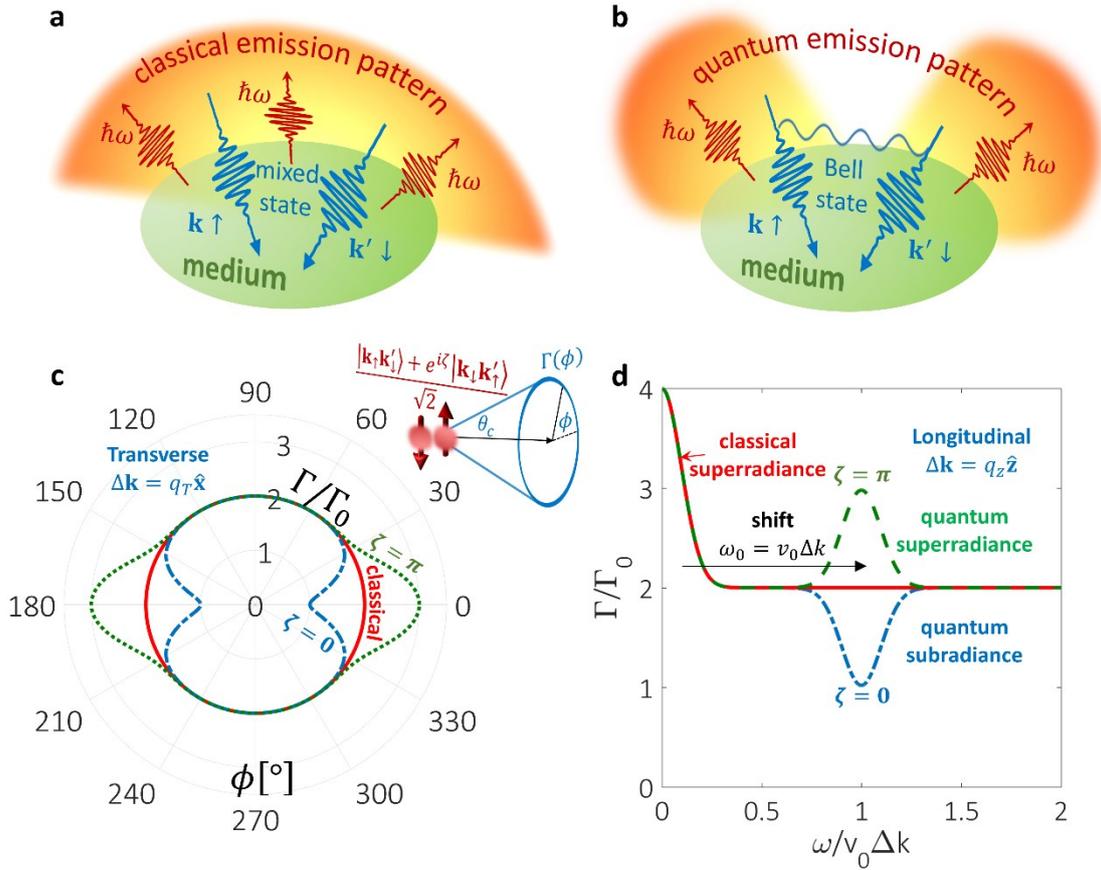

**Figure 2: Shaping light using quantum correlations. (a-b)** Illustration of coherent cathodoluminescence from two correlated particles. In (a), a pair of coincident electrons with different momenta $\mathbf{k}, \mathbf{k}'$ and spins ↑↓ are prepared in a classically correlated (mixed) state and interact with an optical environment – giving classical emission. In (b), the electrons are instead prepared in a path-entangled Bell-state, and the emission pattern is modified, depending explicitly on the phase angle ζ of the electrons' quantum state. (c-d) Quantum shaping of Cherenkov radiation by two-electron Bell-states. **(c)** The rate of Cherenkov photon emission per unit time per unit frequency Γ, normalized by the single-particle rate $\Gamma_0 = \alpha\beta \sin^2\theta_c /2\pi$. The normalized rate $\Gamma/\Gamma_0$ is calculated along the Cherenkov cone and plotted as a function of the azimuthal angle $\phi$ (see inset). For different phases ζ of the electron Bell-state, the radiation pattern is no longer azimuthally symmetric on the cone as in the classical case (red full line), and is either enhanced ($\zeta = \pi$, green dotted line) or suppressed ($\zeta = 0$, blue dashed-dotted line). The enhancement/suppression occurs along an axis defined by the momentum difference $\Delta\mathbf{k}$ of the two-electron Bell-state. The two electron wavepackets that constitute the two-electron Bell-state differ by a transverse wavevector $\Delta\mathbf{k} = \sin\theta_c\, n\omega/c\,\hat{\mathbf{x}} = q_T\hat{\mathbf{x}}$, chosen to match the transverse photon momentum $q_T$. **(d)** Normalized emission rate $\Gamma/\Gamma_0$ vs. normalized frequency in units of $\omega_0 = v_0\Delta k$, where now $\Delta\mathbf{k}$ is chosen parallel to $\hat{\mathbf{z}}$. Near $\omega = v_0\Delta k = \omega_0$, a resonance appears. The magnitude of the spectral feature is governed by the phase angle ζ of the Bell-state, giving quantum super- and subradiance. Electron velocity $v_0 = 0.7c$, refractive index $n = 2$, emitted photon

energy $\hbar\omega = 2$ eV, wavefunction dimensions in (c) $\Delta r_T = 200$ nm and $\Delta z = 1$ nm and in (d) $\Delta r_T = 10$ nm and $\Delta z = 500$ nm, for the transverse and longitudinal sizes, respectively.

**Cherenkov radiation.** As an example of our findings, we consider Cherenkov radiation (CR), observed when a charged particle of velocity $v = \beta c$ surpasses the phase velocity of light in a homogeneous dielectric medium of refractive index $n = n(\omega)$. CR is known to have a broad spectrum, and is characterized by a cone-shaped emission pattern, where the aperture of the cone is determined by the Cherenkov angle $\theta_C = \arccos(1/n\beta)$. For the two-electron cases discussed above, the CR emission rate is found using Eqs. (1-5), yielding

$$\Gamma(\hat{\mathbf{n}}, \omega) = \frac{\alpha\beta}{2\pi}\sin^2\theta\,\delta\left(\cos\theta - \frac{1}{n\beta}\right)\left\{\underbrace{2 + 2\left|\int d^3\mathbf{x}\, e^{-i\frac{n\omega}{c}\hat{\mathbf{n}}\cdot\mathbf{x}}|\varphi(\mathbf{x})|^2\right|^2}_{\text{classical}}\right.$$
$$\left.-\underbrace{\cos\zeta\left[\left|\int d^3\mathbf{x}\, e^{-i\left(\frac{n\omega}{c}\hat{\mathbf{n}}-\Delta\mathbf{k}\right)\cdot\mathbf{x}}|\varphi(\mathbf{x})|^2\right|^2 + \Delta\mathbf{k}\leftrightarrow -\Delta\mathbf{k}\right]}_{\text{quantum}}\right\}, \quad (6)$$

In Eq. (6), note that the first two terms correspond to the classically-correlated state (giving the classical emission rate $\Gamma_c$), while the third term appears only for the quantum-correlated electron Bell-state of phase angle $\zeta$, with $\Gamma_{\Delta\mathbf{k}}$ proportional to the shifted Fourier transform of the wavepacket. The shifted spectrum, together with the phase angle $\zeta$, can be used to tailor quantum super- and subradiant CR.

Fig. 2c-d illustrates this example using two choices of $\Delta\mathbf{k}$: matching the transverse momentum (Fig. 2c) or the longitudinal momentum (Fig. 2d) of the emitted CR photon in a specific optical wavelength. For the quantum-correlated case, this matching results in a shaped emission pattern and spectrum at the chosen wavelength – which does not occur in the classical case (see also the appendix). The quantum phase angle $\zeta$ controls, in the transverse case (Fig. 2c), the quantum suppression and enhancement of the radiation with respect to the classical emission rate in opposite angles on the Cherenkov cone. Similarly, in the longitudinal case (Fig. 2d) the phase $\zeta$ induces super- and subradiance at the chosen resonant frequency $\omega_0$ that is matched to the momentum difference via $\omega_0 = v_0\Delta k$.

**Generalization to $N$ particles.** Our results could be readily applied to the case of many-body states of free charged particles, by the choice of appropriate wavefunctions of the form

$$|\Psi\rangle = \sum_{\{\sigma_1\sigma_2\ldots\sigma_N\}} c_{\{\sigma_1\sigma_2\ldots\sigma_N\}}|\varphi_1\sigma_1, \varphi_2\sigma_2, \ldots, \varphi_N\sigma_N\rangle, \quad (7)$$

where $\sigma_i$ is the spin of particle $i$ occupying the wavepacket $\varphi_i(\mathbf{r})$. Even for the much-higher dimensionality of the wavefunction, the same current correlations of Eq. 2 enable deriving observables as above. When the number of particles grows, entanglement features and Pauli exclusion[50] play an increasingly important role in the shaping of radiation patterns. Looking forward, it is extremely interesting to find which many-body states create macroscopic states of light that widely differ from conventional types of light emission and from classical super- and subradiance. Such findings could serve as concrete evidence for the breakdown of the correspondence principle[51] between Maxwell's equations and quantum electrodynamics.

**Experimental considerations.** Path entangled free electron states were observed in double photoionization from $H_2$ molecules[44,45]. Other directions for entangled electrons have been proposed by exploiting the interactions of free electrons with cavity photons[25,46]. The phase angle of these states could be controlled using path differences[47] or through interaction with an optical field[48]. In addition, spectral modulation of two (or more) entangled electron states can be implemented using photon-induced electron microscopy (PINEM) techniques[49]. We emphasize that the quantum interference exists even in the complete absence of classical interference effects, such as those related to inducing a space charge modulation or time delay between the two electrons. This point is shown in the appendix for any wavepacket shape and for any optical medium. Consequently, our entangled-driven super- and subradiance can enable a clear experimental signature of quantum super- and subradiance from entangled particles.

**Discussion**

We unveiled the role of quantum correlations in enabling a novel form of super- and subradiance from several charged particles. This effect was previously analysed only in the presence of classical correlations. Harnessing the quantum interference between two path-entangled electrons, we showed how the intensity pattern and spectrum of CCL, and generally of any emission process from charged particles, can be selectively enhanced or suppressed, depending on the quantum state of the particles. This capability paves the way towards novel light sources based on collective emission from multi-particle quantum wavefunctions, also suggesting that photoemission by free electrons could serve as a non-destructive quantum measurement of the entanglement between the electrons. Preparation of free-electron entangled states, necessary for realizing such scenarios, is under investigation by different groups[25,46], for example by using a photonic cavity to entangle the electrons.

We first presented the results that led to this work in the CLEO conference in May 2020[52].

**Appendix: Comparison between emission by product states and entangled states**

This section shows that classical super- and subradiance (due to, for example, to a temporal delay between free electrons or a spatio-temporal modulation of the electron charge density) is qualitatively and quantitatively different from the super- and subradiance effects that we find due to entanglement. Let us consider first two electrons prepared in two arbitrary wavepacket modes $\varphi_1(\mathbf{r})$ and $\varphi_2(\mathbf{r})$. We constrain the discussion to electrons distinguishable by spin. We consider two possible cases: a product state

$$|\Psi_{\text{prod}}\rangle = |\varphi_{1\uparrow}\varphi_{2\downarrow}\rangle, \tag{A.1}$$

and an entangled state, of the form discussed in the main text:

$$|\Psi_{\text{ent}}\rangle = \frac{|\varphi_{1\uparrow}\varphi_{2\downarrow}\rangle + e^{i\zeta}|\varphi_{1\downarrow}\varphi_{2\uparrow}\rangle}{\sqrt{2}}, \tag{A.2}$$

The emitted power spectrum can be written as:

$$\frac{d^2P}{d\Omega d\omega} = \frac{\hbar\omega\alpha\beta}{2\pi} \sin^2\theta \, \delta\left(\cos\theta - \frac{1}{n\beta}\right) \Bigg\{ \underbrace{2 + 2\text{Re}\int d^3\mathbf{x} \int d^3\mathbf{x}' \, e^{-i\frac{n\omega}{c}\hat{\mathbf{n}}\cdot(\mathbf{x}-\mathbf{x}')} |\varphi_1(\mathbf{x})|^2 |\varphi_2(\mathbf{x}')|^2}_{\text{product state}}$$

$$- \underbrace{\cos\zeta \int d^3\mathbf{x} \int d^3\mathbf{x}' \, e^{-i\frac{n\omega}{c}\hat{\mathbf{n}}\cdot(\mathbf{x}-\mathbf{x}')} 2\text{Re}\{\varphi_1(\mathbf{x})\varphi_2^*(\mathbf{x})\varphi_2(\mathbf{x}')\varphi_1^*(\mathbf{x}')\}}_{\text{additional entanglement term}} \Bigg\}, \quad \text{(A.3)}$$

where the first two terms encompass all classical super- or subradiance interference effects due to the structure of the chosen wavepackets $\varphi_1(\mathbf{r})$ and $\varphi_2(\mathbf{r})$. The third term is purely quantum and can be nonzero even for cases where no interference related with time delay or density modulation occurs.

**References**

1. Harris, E. G. *A Pedestrian Approach to Quantum Field Theory.* (Dover Publications, 2014).

2. Gross, M. & Haroche, S. Superradiance: An essay on the theory of collective spontaneous emission. *Phys. Rep.* **93**, 301–396 (1982).

3. Dicke, R. H. Coherence in spontaneous radiation processes. *Phys. Rev.* **93**, 99–110 (1954).

4. Guerin, W., Araújo, M. O. & Kaiser, R. Subradiance in a Large Cloud of Cold Atoms. *Phys. Rev. Lett.* **116**, 083601 (2016).

5. Goban, A. *et al.* Superradiance for Atoms Trapped along a Photonic Crystal Waveguide. *Phys. Rev. Lett.* **115**, 063601 (2015).

6. Scheibner, M. *et al.* Superradiance of quantum dots. *Nat. Phys.* **3**, 106–110 (2007).

7. Bienaimé, T., Piovella, N. & Kaiser, R. Controlled Dicke subradiance from a large cloud of two-level systems. *Phys. Rev. Lett.* **108**, 123602 (2012).

8. Jahnke, F. *et al.* Giant photon bunching, superradiant pulse emission and excitation trapping in quantum-dot nanolasers. *Nat. Commun.* **7**, 11540 (2016).

9. Solano, P., Barberis-Blostein, P., Fatemi, F. K., Orozco, L. A. & Rolston, S. L. Super-radiance reveals infinite-range dipole interactions through a nanofiber. *Nat. Commun.* **8**, 1–7 (2017).

10. Gover, A. *et al.* Superradiant and stimulated-superradiant emission of bunched electron beams. *Rev. Mod. Phys.* **91**, (2019).

11. Bostedt, C. *et al.* Linac Coherent Light Source: The first five years. *Rev. Mod. Phys.* **88**, 015007 (2016).

12. Pellegrini, C., Marinelli, A. & Reiche, S. The physics of x-ray free-electron lasers. *Rev. Mod. Phys.* **88**, 015006 (2016).

13. Korbly, S. E., Kesar, A. S., Sirigiri, J. R. & Temkin, R. J. Observation of Frequency-Locked Coherent Terahertz Smith-Purcell Radiation. *Phys. Rev. Lett.* **94**, 054803 (2005).

14. Cook, A. M. *et al.* Observation of Narrow-Band Terahertz Coherent Cherenkov Radiation from a Cylindrical Dielectric-Lined Waveguide. *Phys. Rev. Lett.* **103**, 095003 (2009).

15. O'Shea, P. G. & Freund, H. P. Free-electron lasers: Status and applications. *Science* **292**, 1853–1858 (2001).

16. Ratner, D., Hemsing, E., Gover, † A, Marinelli, A. & Nause, A. Subradiant spontaneous undulator emission through collective suppression of shot noise. *Phys. Rev. Spec. Top. - Accel. Beams* **18**, 050703 (2015).

17. Gover, A., Nause, A., Dyunin, E. & Fedurin, M. Beating the shot-noise limit. *Nat. Phys.* **8**, 877–880 (2012).

18. Ratner, D. & Stupakov, G. Observation of shot noise suppression at optical wavelengths in a relativistic electron beam. *Phys. Rev. Lett.* **109**, 034801 (2012).

19. Remez, R. *et al.* Spectral and spatial shaping of Smith-Purcell radiation. *Phys. Rev. A* **96**, 061801 (2017).

20. Polman, A., Kociak, M. & García de Abajo, F. J. Electron-beam spectroscopy for nanophotonics. *Nat. Mater.* **18**, 1158–1171 (2019).

21. Shentcis, M. *et al.* Tunable free-electron X-ray radiation from van der Waals materials. *Nat. Photonics* **14**, 686–692 (2020).

22. Adamo, G. *et al.* Light well: A tunable free-electron light source on a chip. *Phys. Rev. Lett.* **103**, 113901 (2009).

23. Pan, Y. & Gover, A. Spontaneous and stimulated emissions of a preformed quantum free-electron wave function. *Phys. Rev. A* **99**, 052107 (2019).

24. Di Giulio, V., Kociak, M. & de Abajo, F. J. G. Probing quantum optical excitations with fast electrons. *Optica* **6**, 1524 (2019).

25. Kfir, O. Entanglements of Electrons and Cavity Photons in the Strong-Coupling Regime. *Phys. Rev. Lett.* **123**, 103602 (2019).

26. Bendaña, X., Polman, A. & García De Abajo, F. J. Single-photon generation by electron beams. *Nano Lett.* **11**, 5099–5103 (2011).

27. Gover, A. & Yariv, A. Free-Electron-Bound-Electron Resonant Interaction. *Phys. Rev. Lett.* **124**, 064801 (2020).

28. Zhao, Z., Sun, X.-Q. & Fan, S. Quantum entanglement and modulation enhancement of free-electron-bound-electron interaction. *arXiv:2010.11396* (2020).

29. Hirschmugl, C. J., Sagurton, M. & Williams, G. P. Multiparticle coherence calculations for synchrotron-radiation emission. *Phys. Rev. A* **44**, 1316–1320 (1991).

30. Saleh, B. E. A. & Teich, M. C. *Fundamentals of photonics*. (Wiley-Interscience, 2007).

31. García De Abajo, F. J. & Giulio, V. Di. Quantum and Classical Effects in Sample Excitations by Electron Beams. *arXiv:2010.13510* (2020).

32. Angioi, A. & Piazza, A. Di. Quantum Limitation to the Coherent Emission of Accelerated Charges. *Phys. Rev. Lett.* **121**, (2018).

33. Kling, P., Giese, E., Carmesin, C. M., Sauerbrey, R. & Schleich, W. P. High-gain quantum free-electron laser: Emergence and exponential gain. *Phys. Rev. A* **99**, 053823 (2019).

34. Peter Kling, Enno Giese, Rainer Endrich, Paul Preiss, Roland Sauerbrey, W. P. S. What

defines the quantum regime of the free-electron laser? *New J. Phys.* **17**, 123019 (2015).

35. Robb, G. R. M. & Bonifacio, R. Coherent and spontaneous emission in the quantum free electron laser. *Phys. Plasmas* **19**, 073101 (2012).

36. Kaminer, I. *et al.* Quantum Čerenkov Radiation: Spectral Cutoffs and the Role of Spin and Orbital Angular Momentum. *Phys. Rev. X* **6**, 011006 (2016).

37. García de Abajo, F. J. Optical excitations in electron microscopy. *Rev. Mod. Phys.* **82**, 209–275 (2010).

38. Remez, R. *et al.* Observing the Quantum Wave Nature of Free Electrons through Spontaneous Emission. *Phys. Rev. Lett.* **123**, 060401 (2019).

39. Hayun, A. Ben *et al.* Shaping Quantum Photonic States Using Free Electrons. *arXiv:2011.01315* (2020).

40. Novotny, L. & Hecht, B. *Principles of nano-optics*. *Principles of Nano-Optics* (Cambridge University Press, 2006).

41. Scheel, S. & Buhmann, S. Y. Macroscopic QED - concepts and applications. *Acta Phys. Slovaca* **58**, 675 (2008).

42. Rivera, N. & Kaminer, I. Light–matter interactions with photonic quasiparticles. *Nature Reviews Physics* **2**, 538–561 (2020).

43. Charles Roques-Carmes, Steven E. Kooi, Yi Yang, Aviram Massuda, Phillip D. Keathley, Aun Zaidi, Yujia Yang, John D. Joannopoulos, Karl K. Berggren, I. K. & M. S. Towards integrated tunable all-silicon free-electron light sources. *Nat. Commun.* **10**, 3176 (2019).

44. Akoury, D. *et al.* The simplest double slit: Interference and entanglement in double photoionization of H2. *Science (80-. ).* **318**, 949–952 (2007).

45. Waitz, M. *et al.* Two-Particle Interference of Electron Pairs on a Molecular Level. *Phys. Rev. Lett.* **117**, 083002 (2016).

46. Mechel, C. *et al.* Imaging the collapse of electron wave-functions: the relation to plasmonic losses. in *Conference on Lasers and Electro-Optics (2019), paper FF3M.6* FF3M.6 (The Optical Society, 2019).

47. Lichte, H. & Lehmann, M. Electron holography—basics and applications. *Reports Prog. Phys.* **71**, 016102 (2008).

48. Echternkamp, K. E., Feist, A., Schäfer, S. & Ropers, C. Ramsey-type phase control of free-electron beams. *Nat. Phys.* **12**, 1000–1004 (2016).

49. Priebe, K. E. *et al.* Attosecond electron pulse trains and quantum state reconstruction in ultrafast transmission electron microscopy. *Nat. Photonics* **11**, 793–797 (2017).

50. Rom, T. *et al.* Free fermion antibunching in a degenerate atomic Fermi gas released from an optical lattice. *Nature* **444**, 733–736 (2006).

51. Bohr, N. & Nielsen, J. R. *The correspondence principle : 1918-1923*. (North-Holland Pub. Co, 1976).

52. Karnieli, A., Rivera, N., Arie, A. & Kaminer, I. Coneference talk: Unveiling Emitter Wavefunction Size via the Quantum Coherence of its Radiation. *Conference on Lasers*

*and Electro-Optics (May 2020)* FTu3D.5 Available at: https://osa.zoom.us/rec/play/YNzAU7COMEQOWVKHnYEFqriGAy_kaV7kCHoiRh22NXLxtZfpdwRI3F3o8GRvQFR_JCBg0nWwkaMjFLnk.KUdmeKsaWritaj9d?continueMode=true&_x_zm_rtaid=5qoJm4GnRPmk0y-XPZPKNQ.1603880501651.f698ab76cabac4272ad6465ec6fc32ce&_x_zm_rhtaid=259.